\documentclass[fleqn,12pt,twoside]{article}
\usepackage{espcrc1}
\usepackage{myespcrc1_changes}
\usepackage{graphicx}
\usepackage{geometry}
\usepackage{amssymb}
\begin{document}
\newcommand{\PRB}{Phys. Rev. B }
\newcommand{\PRL}{Phys. Rev. Lett. }
\renewcommand{\vec}[1]{ {\mathbf #1 } }
\newcommand{\e}{{\mathrm{e}}}
\renewcommand{\rho}{\varrho}
\renewcommand{\phi}{\varphi}
\renewcommand{\epsilon}{\varepsilon}
\renewcommand{\i}{{\mathrm{i}}}
\newcommand{\diff}{{\mathrm{d}}}
\setlength{\mathindent}{3em}
\pagestyle{empty}
\hyphenation{Gra-du-ier-ten-kolleg}

\title{\vspace*{-2.7cm} \center Ground state energies of quantum dots in high 
magnetic fields: \\ A new approach}

\author{\center J. Kainz\thanks{Corresponding author; email: Josef.Kainz@physik.uni-regensburg.de}, 
S. A. Mikhailov\thanks{Present address: Institut 
f\"ur Physik, Universit\"at Augsburg, D-86135 Augsburg, Germany.}, 
A. Wensauer, U. R\"ossler \\
Institut f\"ur Theoretische Physik, Universit\"at Regensburg,
\\ 
D-93040 Regensburg, Germany
}

\maketitle

\begin{abstract}
\begin{center} 
\bf Abstract
\end{center} 

\noindent
We present a new method for calculating ground state properties of quantum dots in high magnetic fields.
It takes into account the equilibrium positions of electrons in a Wigner cluster to minimize the interaction
energy in the high field limit. Assuming perfect spin alignment the many-body trial function is a 
single Slater determinant of overlapping oscillator functions from the lowest Landau level centered at and near 
the classical equilibrium positions. 
We obtain an analytic expression for the ground state energy 
and present numerical results for up to $N=40$. 
\end{abstract}

\section{Introduction}
\label{introduction}

Quantum dots, often referred to as artificial atoms \cite{kastner}, were intensively studied in recent years 
both experimentally \cite{Ashoori96,tarucha} and theoretically \cite{hawrylak,yang,reimann2000,maksym,%
bolton,Egger99,harju,pederiva,filinov,pfannkuche,chamon,reusch,hirose,steffens,wensauer}. 
A number of different theoretical methods, including, e.g., 
exact diagonalization \cite{hawrylak,yang,reimann2000,maksym}, 
quantum Monte Carlo (QMC) \cite{bolton,Egger99,harju,pederiva,filinov}, 
Hartree-Fock \cite{pfannkuche,chamon,reusch} 
and density-functional calculations \cite{hirose,steffens,wensauer}, 
were employed for investigating the ground state properties of dots, which are responsible for 
many of the experimentally observed effects.

In this paper we present a new concept to calculate ground state properties of quantum dots in high magnetic 
fields, with application to systems with up to 40 electrons. 
This approach is motivated by the evidence that in high magnetic fields $B$ quantum 
dot electrons tend to form a Wigner cluster \cite{bolton,maksym}.
Therefore, we construct a ground-state many-body variational wave function 
in the form of a Slater determinant with (overlapping) single-particle wave 
functions centered at or near equilibrium positions of classical point 
particles.
The classical configurations are calculated either analytically (for $N\le 8$) or using 
the Monte Carlo minimization \cite{bolton,bedanov}. 
Expectation values of observables can be formulated analytically. Here we
present ground state energies for disc-like quantum dots with up to 40
electrons.

\section{The model Hamiltonian}
\label{modelhamiltonian}

We consider a system of $N$ electrons moving in the plane $z=0$, in the lateral potential $ V({\mathbf r }) = m^* \omega_0^2 (x^2+y^2)/2 $, and in the perpendicular magnetic field ${\mathbf B }=(0,0,B)$. The Hamiltonian of the system has the form

\begin{equation}
\hat{H} = \sum_{i=1}^N \left[ \frac{ [{\mathbf p}_i + e {\mathbf A}({\mathbf r}_i)]^2}{2m^*} + 
V(\vec{r}_i) \right] + V_C +E_Z,
\label{hamiltonian}
\end{equation}
where 
$\vec{A} (\vec{r}) = (B/2) (-y,x) $, $ V_C $ 
is the Coulomb energy of electron-electron 
interaction and $E_Z$ is the Zeeman energy. The solutions of the single-particle problem (Fock and Darwin, \cite{fockdarwin}) are 
given by
$ \psi_{n,l}(r, \phi) \propto  \e^{\i l \phi} r^{|l|} e^{- r^2/2}  L_n^{|l|}(r^2) $,
where $r^2=(x^2 + y^2)/ \lambda^2 $, 
$ \lambda^2 = (\hbar / m^*) [\omega_0^2 + ( \omega_c / 2 )^2 ]^{-1/2} $,  $ \omega_c = eB/m^* $ 
is the cyclotron 
frequency, $L_n^{|l|}$ are associated Laguerre polynomials, $n\in {\mathbb{N}}_0$  is the number of nodes, and $l\in \mathbb{Z}$ is the angular
momentum quantum number.

\section{The new approach}
\label{newapproach}

In strong magnetic fields electrons tend to localize around the classical equilibrium positions, and their spins 
to be aligned along the magnetic field. 
This tendency suggests to write the trial many-body wave function 
$\Psi_L$ in the form of a Slater determinant (as has been done in Ref. 
\cite{mikhailov} for a related problem) 

\begin{equation}
\label{Psi_L}
\Psi_L = \det| \chi^{(L)}_{ij}|/\sqrt{N!}
, \; 
\chi^{(L)}_{ij}=\psi_{0,-L}(\vec{r}_i - \vec{R}_j) \exp{ \left( - \frac{\i e}{2 \hbar } 
\left( \vec{B} \times \vec{R}_j \right) \cdot
 \vec{r}_i 
 \right) }
,
\end{equation}
where $ \chi^{(L)}_{ij} $
are Fock-Darwin wave functions with an additional phase factor caused by 
a magnetic translation \cite{landau} of the $i$-th 
electron to the center at $ \vec{R}_j $. 
The non-negative integer $L$ and the vectors $ \vec{R}_1, \ldots , \vec{R}_N $ are  free parameters of the theory. In general, 
the single particle wave functions $\chi^{(L)}_{ij}$ are not orthogonal, and the overlap of the neighbor single-particle states 
needs not to be small. 

Using the trial wave functions $ \Psi_L $ we analytically  calculate expectation values of the energy and the density of electrons. 
The energy is minimized with respect to the variational parameters $L$ and $ \vec{R}_1, \ldots , \vec{R}_N $.

\section{Results}
\label{results}

\begin{figure}[ht!]
\mbox{\includegraphics[width = 9 cm ]{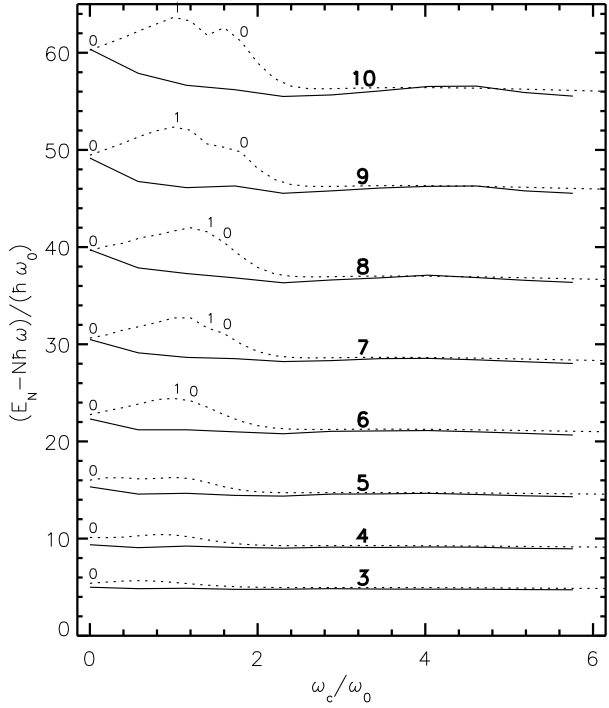}}
\mbox{\parbox[b]{6.5cm}{
\caption{
\label{fig_bolton}
Calculated expectation values for the ground state energy (dotted) of $N=3-10$ electrons. Bold numbers designate the 
number of electrons, 
$ \omega = \left( \omega_0^2 + \omega_c^2 / 4 \right)^{1/2} $. Full curves show the QMC results from 
Ref. \cite{bolton}.
For this plot we have chosen $l_0/a_B=1.91$, where $l_0^2=\hbar/m^*\omega_0$ and $a_B$ is the effective Bohr radius. 
The Zeeman energy with effective g-factor $ g^* = -0.44 $ was taken into account.
\newline
\vspace*{0.5 cm}
}
}
}
\vspace*{-1.6cm}
\end{figure}

Figure \ref{fig_bolton} shows the calculated ground state energy of a system of 3--10 electrons, together with 
reference data from Ref. \cite{bolton}, obtained by the QMC method. The small numbers 0 and 1 over the dotted lines indicate 
that, starting from the corresponding points the trial wave functions $\Psi_{L=0}$ or $\Psi_{L=1}$ gave the lowest energy. 
The parameters ${\bf R}_i$ were chosen according to classical equilibrium configurations and were not varied 
in Figure \ref{fig_bolton}. 

As seen from the Figure, our results are very close to (and sometimes even lower than) the QMC results at  $ \omega_c / \omega_0 \gtrsim 2 $. In this range typical relative deviations of the total energy are $ 2 \% $ and less. Optimizing the parameters ${\bf R}_i$ we could get even better agreement with the QMC results in the range $ 0.6 \lesssim \omega_c / \omega_0 \lesssim 2.4 $ (the optimized radius of electron shells is up to 20\% larger at small magnetic fields than the classical radius). 
The accuracy of our results shows a tendency to increase with larger $B$ and/or $N$. 
At the upper range of the magnetic field the relative deviation is about
$ 1 \% $ for $N \le 6 $ and even less for higher $N$.
This remarkably good agreement and its tendency to become even better with 
increasing electron number demonstrates the capability of the new method and is
promise for obtaining reliable results for even higher electron numbers.
\begin{figure}[ht!]
\mbox{\includegraphics[width = 9cm ]{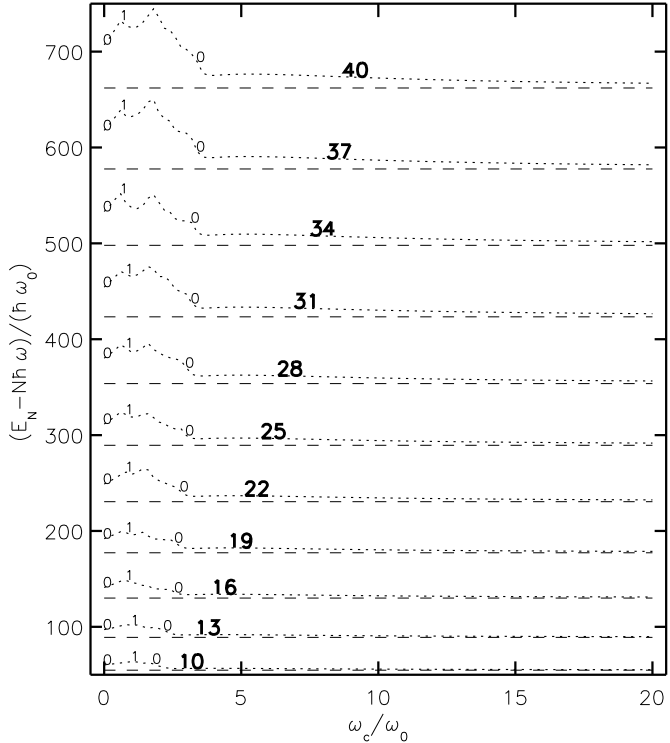}}
\mbox{
\parbox[b]{6.5cm}{
\caption{
\label{fig_qm_class}
Calculated ground state energies for 10--40 electrons (dotted curves).
The horizontal dashed lines indicate the value of classical ground state energy $E_N^{class}/\hbar\omega_0$.
The meaning of the numbers in the plot and the parameter $l_0/a_B$ are as in figure \ref{fig_bolton}, except $g^*=0$ here.
\vspace*{1.5cm}
}
}
}
\vspace*{-1.3cm}
\end{figure}

Figure \ref{fig_qm_class} gives calculated ground state energies $ E_N $ in the
range of up to $ N=40 $,
for which suitable exact reference values were not available. 
With increasing magnetic field we find a clear convergence of 
$ E_N - N \hbar \omega $ towards $ E_N^{class} $,
so that we can write
\begin{equation}
E_N = E_N^{class} + N \hbar \sqrt{ \omega_0^2 + \left( \omega_c/2 \right)^2 }
+\delta E_N.
\label{approx}
\end{equation}
The correction $\delta E_N$ is positive, tends to zero at $B\to\infty$, and for $ 10 \le N \le 40 $ it is  
smaller than 2\% at $\omega_c/\omega_0> 6 $ and smaller than 0.6\% at $ \omega_c/\omega_0 > 18 $. 
Equation (\ref{approx}) refines the rigorous mathematical result by Yngvason \cite{yngvason} and has a 
transparent physical meaning: The quantum mechanical ground state energy of the dot is a sum of the 
classical potential energy, the quantum mechanical kinetic energy, and the small rest energy $\delta E_N$, which
decreases \mbox{with $B$}.

\section{Conclusion}
\label{conclusion}

We have presented a new method to calculate ground state expectation values 
for parabolic quantum dots in (high) magnetic fields. The results 
were obtained by evaluating the expectation value, for which we found an 
analytical expression 
using a specially designed trial wave function. It takes into 
account the
classical equilibrium positions of the electrons and has the correct symmetry 
for a fermion system with fully polarized spins.
In the high magnetic field regime our results compare favourably with QMC data,
but are much less demanding in computer time. Therefore, an extension to higher
electron numbers becomes possible as we demonstrate by showing results for up
to 40 electrons.
We put forward a simple interpretation of our results for the ground state energy, Eq. (\ref{approx}), and gave 
the upper estimate for the rest energy term $\delta E_N$ in the studied range of $N$ and the magnetic field.

This work was supported by the Deutsche Forschungsgemeinschaft within
the Graduiertenkolleg (GK 176) {\sl Komplexit\"at in Festk\"orpern}.

\end{document}